\documentclass[12pt,preprint]{aastex}
\bibpunct{(}{)}{;}{a}{}{,} % to follow the A&A style

\shorttitle{Period Evolution of AXPs in the Context of Quark Stars}
\shortauthors{Niebergal et al.}

\begin{document}

\title{Magnetic Field Decay and Period Evolution of Anomalous X-Ray Pulsars in the Context of Quark Stars }

\author{Brian Niebergal, Rachid Ouyed and Denis Leahy }

\affil{Department of Physics and Astronomy, University of Calgary, 
2500 University Drive NW, Calgary, Alberta, T2N 1N4 Canada}

\email{bnieber@iras.ucalgary.ca}

%\abstract{}{Aims}{Methods}{Results}{}
\begin{abstract}
We discuss a model wherein soft gamma-ray repeaters (SGRs), 
anomalous X-ray pulsars (AXPs), and radio quiet isolated neutron stars (RQINSs)
are all compact objects exhibiting superconductivity, namely color-flavor 
locked quark stars. 
In particular we calculate the magnetic field decay due to the
expulsion of spin-induced vortices from the star's superfluid-superconducting interior,
and the resultant spin-down rate.  
We find that, for initial parameters characteristic of AXPs/SGRs 
($10^{13}<B<10^{14}~\rm{G}$; $3<P<12~\rm{s}$), 
the magnetic field strengths and periods
remain unchanged within a factor of two
for timescales of the order of $5\times 10^{5} - 5\times 10^{7}~\rm{yrs}$
given a quark star of radius $10~\rm{km}$.
Within these timescales, we show that the observed period clustering in RQINSs
can be explained by compactness, as well as calculate how
the magnetic field and period evolve
in a manner concurrent with RQINS observations.

\end{abstract}

%\maketitle

\section{Introduction}

Soft $\gamma$-ray repeaters (SGRs) are sources of recurrent, 
short ($t \sim 0.1\,\mathrm{s}$), intense ($L \sim 10^{44}~\rm{ergs}$) 
bursts of $\gamma$-ray emission with a soft energy spectrum.
The normal pattern of SGRs are intense activity periods 
which can last weeks or months, separated by quiescent phases 
lasting years or decades. 
%The five known SGRs are located in our Galaxy or, in the case of SGR 0526-66, 
%in the Large Magellanic Cloud.  
The three most intense SGR bursts ever recorded 
were the 5 March 1979 giant flare of SGR 0526-66 
(Mazets et al.\ 1979), the similar 28 August 1998 giant 
flare of SGR 1900+14 and the 27 December 2004 burst (SGR 1806-20).  
AXPs are similar in nature but with a somewhat weaker intensity and
no recurrent bursting.
Several SGRs/AXPs have been found to be X-ray pulsars with an 
unusually high spin-down rate of $\dot{P} / P \sim 10^{-10}$~s$^{-1}$, 
usually attributed to magnetic braking caused by a super-strong 
magnetic field. 

The model normally reserved for SGRs/AXPs is the magnetar model, however
it has been suggested (Ouyed et al.~2004) that CFL (color-flavor 
locked) quark stars could be also responsible for their activity.  
In this quark star model, we assume a neutron star has made the
transition from hadronic to superfluid-superconducting
CFL quark matter. 
Through the Meissner effect,
the quark star's interior magnetic field is forced 
inside rotationally induced vortices that are aligned with the rotation
axis of the star. 
The exterior dipole field is forced to align with the rotation axis
as simulated in Ouyed et al.(2005), with application to SGRs/AXPs.

In this Letter, we extend the model by studying the post-alignment spin-down phase.
Since the star's interior is a superconducting-superfluid, the 
number of vortices contained within has a quantized relation to the spin period.
As the star spins down, the vortices containing the magnetic field
are forced to the surface where they are expelled. 
The contained magnetic field decays through 
reconnection events, thus lowering the spin-down rate 
and providing X-ray emission.
Our model bears some similarities to the picture presented in Srinivasan et al.
(1990) for the case of neutron stars where vortex expulsion is intimately
related to the rotation history of the neutron star. 
What makes our model work is the the fact that the 
fluxoids are trapped inside the vortices, and the absence of a crust 
(absence of electrons) which removes any pinning forces.

Using this model of vortex expulsion, along with simple magnetic dipolar braking,
we calculate the period evolution and magnetic field decay of the quark star.
We also show how the evolution of quark stars (AXPs/SGRs) leads to parameters
indicative of radio quiet isolated neutron stars (RQINSs; see \S~\ref{sec:RQINSs}). 
We then show how the compactness of a quark star governs 
the period clustering in RQINSs.

The paper is presented as follows:
In \S~\ref{sec:qm_formation} we describe the formation of quark stars
and how the Meissner effect constrains the
magnetic field into vortices aligned with the rotation axis.
We then calculate the magnetic field decay and period evolution
during the quiescent phase in \S~\ref{sec:field_decay}. 
Then, in \S~\ref{sec:RQINSs}, we show how our model
predicts magnetic field strengths, periods, and ages consistent with
RQINS observations, and, discuss the results
in the context of a $P-\dot{P}$ diagram. Finally in \S~\ref{sec:RQINSs},
we describe the relation between the compactness of the quark star and
the resultant period clustering.  We conclude in \S~\ref{sec:conclusion}.

%%%%%%%%%%%%%%%%%%%%%%%%%%%%%%%%%%%%%%%%%%%%%%%%%%%%%%%%%%%%%%%%%

\section{Foundation of the Model}\label{sec:qm_formation}

Assume a quark star is born with a temperature $T > T_{\rm c}$
($T_{\rm c}$ is the critical temperature below which superconductivity sets in),
and enters a superconducting-superfluid phase in the core as
it cools by neutrino emission (Ouyed et al. 2002; Ker\"anen et al. 2005), 
and contracts due to spin-down.
The front quickly expands to the entire star followed
by the formation of rotationally induced vortices,
analogous to rotating superfluid $^3$He (the vortex lines
are parallel to the rotation axis; Tilley\&Tilley 1990).
Via the Meissner effect, the magnetic field is partially screened
from the regions outside the vortex cores.
Now the system will consist of, alternating
regions of superconducting material with a screened magnetic
field, and the vortices where most of the magnetic field resides.
As discussed in Ouyed et al. (2004), this has interesting consequences
on how the surface magnetic field adjusts
to the interior field which is confined in the vortices.
In Ouyed et al. (2005) we performed numerical simulations 
of the alignment of a quark star's exterior field, and, found that the physics 
involved was indicative of SGR activity\footnote{See simulations: 
www.capca.ucalgary.ca/\~{}bniebergal/meissner/}.

It has been shown that pure CFL matter is rigorously electrically neutral  
despite the unequal quark masses (Rajagopal \& Wilczek 2001).
However other work (Usov 2004; and references therein)
indicates a thin crust ($M_{\rm crust, max} = 10^{-5}M_{\odot}$) 
is allowed around a quark star due to 
surface depletion of strange quarks.  
In our model we have assumed no depletion of strange quarks 
which implies a bare quark star.
Another simplicity of our model resides in the fact that we have
a single superconducting fluid (the CFL phase). In the case of neutron stars
(e.g. Konenkov \& Geppert 2000),
one has to deal with the neutron superfluid inducing the neutron vortices parallel
to the rotation axis and the proton superconductor inducing the fluxoids 
(the magnetic field concentrated into quantized proton vortex lines) in a direction
parallel to the magnetic field.  In our case, having a single superconducting-superfluid
implies that fluxoids are contained inside the vortex cores.

Thus, the forces at play are quite different for the CFL quark star than for a 
neutron star: 
i) the drag force induced by electron scattering is non-existent in our model 
since no electrons are admitted in pure CFL matter; ii)
in the case of neutron stars, there exists a 
force on the vortices due to the variation of the neutron or proton superfluid gaps 
($\Delta$) with density which can expel or trap the vortices (Hsu 1999). 
In quark matter, the variation of the gap is not well constrained and 
it is common to assume the BCS relation 
$\Delta \propto \sqrt{1- (T/T_c)^2}$. The nearly uniform
temperature implies a nearly uniform gap inside the quark star, thus, no vortex trapping;
iii) the absence of a crust removes all possible surface pinning
of the vortices and the fluxoids (this also implies no Magnus force).
Even in a thin crust case 
the superconducting matter and thus the vortices and fluxoids 
do not extend into the crust, which is suspended $\sim100$-$1000$ fermi 
above the surface of the star (Alcock et al. 1986).   
We argue that crustal pinning can be neglected in this case too.
iv) The remaining force is the buoyancy force, which is not counteracted 
by any pinning forces. Thus the spin-down determines the rate of vortex expulsion.

%%%%%%%%%%%%%%%%%%%%%%%%%%%%%%%%%%%%%%%%%%%%%%%%%%%%%%%%%%%%%%%%%

\section{ Magnetic Field Decay and Spin Evolution }\label{sec:field_decay}

Following the initial alignment event, is the quiescent  
spin-down phase where the outermost vortices are pushed to the surface 
and expelled (Ruutu et al. 1997). 
The magnetic field contained within the vortices is also expelled and
annihilates through magnetic reconnection events near the surface of the star
causing energy release presumably in the X-ray regime.
The number of
vortices decreases slowly with spin-down leading to continuous, quiescent
energy release which can last until
the magnetic field is insufficiently strong to produce detectable emission.

In the aligned-rotator model the star spins-down by
magnetospheric currents escaping through the light cylinder.
 For a neutron star, these currents originate in the crust.
 Instead, in our model, pair production from  magnetic reconnection\footnote{Since our model does not allow for a crust or electrons,
  the parallel electric field induced by the potential drop along the
  magnetic field lines (as happens in a neutron star;  e.g.~M\'esz\'aros 1992) 
  is insufficient to remove charges from the quark star. Instead
   the current necessary for the braking torque is provided by the  $e^{+}e^{-}$
    pairs generated by magnetic reconnection events. First estimates
     show that for the magnetic fields in our model, pair production
     provides sufficient current to remove angular momentum from the star.}
  supplies the currents.   The approximation for an aligned rotator Is 
(i.e.~see M\'esz\'aros 1992),
\begin{equation}\label{eqn:spindown}
\frac{\dot{\Omega}}{\Omega} \approx -\frac{B_{\rm{QS}}^2 R_{\rm{QS}}^6 \Omega^2}{I c^3} ,
\end{equation}
where $I$ is the quark star's moment of inertia.
The vortex annihilation rate, given in Ouyed et al. (2004; Eq.~22),
and Eq.~\ref{eqn:spindown} are then solved simultaneously 
to give the magnetic field decay,
\begin{equation}\label{eqn:bdecay}
B_{\rm{QS}}\left(t\right) = B_{\rm{QS,}0}
  \left[1 + \frac{t}{\tau}\right]^{-\frac{1}{6}} ,
\end{equation}
and period evolution,
\begin{equation}\label{eqn:pdecay}
P_{\rm{QS}}\left(t\right) = P_{\rm{QS,}0}
  \left[1 + \frac{t}{\tau}\right]^{\frac{1}{3}} .
\end{equation}
Where $B_{\rm{QS,}0}$ is the initial magnetic field strength
immediately after the burst at the surface of the quark star, 
$P_{\rm{QS,}0}$ is the initial period, and 
$R_{10,\rm{QS}}$ is the quark star radius in units of $10~\rm{km}$.
We have also defined a characteristic age (in years) of,
\begin{equation}
\tau_{yr} = 5\times 10^4 \\
  \left(\frac{10^{14}~\rm{G}}{B_{\rm{QS,}0}}\right)^2
  \left(\frac{P_{\rm{QS,}0}}{5~\rm{s}}\right)^2
  \left(\frac{M_{\rm{QS}}}{M_{\odot}}\right)
  \left(\frac{10~\rm{km}}{R_{\rm{QS}}}\right)^4.
\end{equation}

Eqs.~(\ref{eqn:bdecay} \& \ref{eqn:pdecay}) 
naturally produce the field decay behavior sought by
Colpi et al. (1999), who used
a phenomenological power law to describe different avenues of magnetic field decay.
However, because we solve period and magnetic field simultaneously,
our model predicts a different period evolution.
Colpi et al. (1999) also argue that some efficient mechanism 
of magnetic field expulsion from the star's interior must exist
in order to explain a short field decay timescale.
Our model provides a natural explanation since the absence of a crust
allows efficient expulsion of unpinned vortices.

In our case,
as illustrated in Figure~\ref{fig:b_decay},
both the period and magnetic field remain unchanged 
to within a factor of two for 
$5\times 10^5$ to $5\times 10^7~\rm{yrs}$ given a quark star radius of $10~\rm{km}$ and 
field strengths typical of AXPs/SGRs in the quark star model 
($10^{13}$ - $10^{14}~\rm{G}$). 
Therefore we have shown that, if we presume long-term period evolution
is dictated by magnetic field decay, 
then the periods of AXPs/SGRs will remain close to their initial periods 
for timescales relevant to observations.  
Thus, our model may provide answers to issues raised 
by Psaltis \& Miller (2002), and, observations of variable braking indices
in AXPs/SGRs made by Kaspi et al.~(2000).  Variable braking indices in our model
would most likely be due to an unsteady magnetic field decay which
is beyond the scope of this work.
%, however, we are currently working on numerical simulations to model this process.
Psaltis \& Miller (2002) also argue that the final periods
of AXPs/SGRs will be no greater than $\sim 12~\rm{s}$, thus, AXPs/SGRs
that are born in a narrow range of periods should remain so indefinitely.
However, they omitted magnetic field decay in their stochastic calculations. 
Our results, based on field decay, indicate that the final periods may indeed 
be larger but only significantly after timescales of 
$5\times 10^{5}$ to $5\times 10^{7}$ years
(for $10^{13}<B<10^{14}~\rm{G}$; see Figure~\ref{fig:b_decay}).
However at these timescales, the likelihood
of observing objects with periods greater than $12~\rm{s}$ is 
low because of the significant decay in field strength, and 
dependence of luminosity on $B^4$ (Ouyed et al. 2004; Eq.~24).
Consequently our model predicts that, since AXPs/SGRs are born within a narrow
range of periods, they will remain so for at least $10^6~\rm{yrs}$ depending
on how compact the star is, after which the range will broaden slightly over time.
This will be discussed in detail in Section~\ref{sec:compactness}.

%%%%%%%%%%%%%%%%%%%%%%%%%%%%%%%%%%%%%%%%%%%%%%%%%%%%%%%%%%%%%%%%%

\section{ Radio Quiet Isolated Neutron Stars }\label{sec:RQINSs}

Radio quiet isolated neutron stars (RQINSs) 
are a class of older ($\sim 10^{6}~\rm{yrs}$) 
stars possessing strong magnetic field strengths ($10^{13}$ to $10^{14}~\rm{G}$) 
and exhibit a clustering in their observed periods similar to that
of AXPs and SGRs.
RQINSs have previously been speculated to be related to AXPs and SGRs 
(see Treves et al. 2000 for a review), however, using our model
we will describe how RQINSs are a natural consequence of 
the magnetic field decay due to vortex expulsion in quark stars.

Firstly RQINSs exhibit no radio pulsations, which in our model 
is a necessary consequence of the AXP/SGR burst which causes the magnetic 
field to align with the star's rotation axis.
Furthermore after the quark star's field has aligned, 
it will spin-down through magnetic braking as described 
in Eqs.~\ref{eqn:bdecay} \& \ref{eqn:pdecay},
and, for ages of the order of $\sim 10^{6}~\rm{yrs}$, 
we arrive at results indicative of RQINSs. 
Specifically, if a quark star experiencing an AXP/SGR burst is 
born with a period of $P_0 = 5~\rm{s}$ and magnetic field
strength of $B_0 = 10^{14}~\rm{G}$, 
then by the time it reaches ages estimated for RQINSs it will
have attained a period of $11~\rm{s}$ and it's field will
have decayed to $\sim 5\times 10^{13}~\rm{G}$. 
This is illustrated in Figure~\ref{fig:b_decay}.  Here, the decrease
in field strength by a factor of two results in a decrease in luminosity
by a factor of $2^4$ (see Ouyed et al. 2004; Eq.~24), which suggests
that only RQINSs possessing an initially strong field are 
more likely to be detectable.

\subsection{ Evolution in the $P$-$\dot{P}$ Diagram}
Figure~\ref{fig:p_pdot} describes the period evolution of a recently born quark
star on a $P-\dot{P}$ diagram for various initial surface magnetic field strengths.
The period derivative, in our case of a magnetic field decaying through vortex expulsion,
is attained from Eq.~\ref{eqn:pdecay}.
The parameters selected for the quark star in Figure~\ref{fig:p_pdot} are, 
a mass of $1M_{\odot}$, 
radius of $10~\rm{km}$, and initial period of $5~\rm{s}$.
Increasing the radius will shift the evolutionary tracks 
upwards, whereas changing the mass has little effect, 
and, selecting different initial periods shifts the tracks left or right. 
So, with expected quark star parameters, all RQINSs
follow evolutionary tracks backwards to the region in the 
$P$-$\dot{P}$ diagram suggestive of AXPs/SGRs. 
More specifically, Figure~\ref{fig:p_pdot} shows that
the source indicated by the small square (RXJ0720.4-3125) 
is $\sim 5\times10^{5}~\rm{yrs}$ old and has an initial 
field strength of $\sim 10^{14}~\rm{G}$, given the
observed period of $8.391~\rm{s}$ and period derivative of 
$\sim 1.5\times 10^{-13}\rm{s~s}^{-1}$.

Figure~\ref{fig:p_pdot} also shows two RQINSs which
are marked by down-arrows, indicating that only upper-limits 
on $\dot{P}$ are known (Pons et al. 2005). In the context of quark stars,
this translates into 
upper-limits on the field strengths of the order of $10^{15}~\rm{G}$, 
and a lower-limit on their age of $\sim 10^{4}~\rm{yrs}$.

\section{ Compactness and Period Clustering }\label{sec:compactness}

Although there may be an insufficient number of observed RQINSs
to conclude definitely the exact range of periods they are clustered into,
Pons et al. (2005) shows that all the periods of observed RQINSs so far
are between the same range as AXPs and SGRs ($3$-$12~\rm{s}$).  
This concurs with our results in that,
after $10^6~\rm{yrs}$, only highly compact stars will have periods which
do not deviate much from the range of their progenitor AXPs/SGRs 
(see Figure~\ref{fig:p_cluster}).  The standard neutron star model
for AXPs/SGRs spinning down due only to dipole radiation has 
$\dot{P} \propto B^2 R^4$, which negates the possibility of period clustering
after $10^6~\rm{yrs}$. In our model, the magnetic field is expelled from the
star's interior and so decays in time, which in turn decreases the spin-down rate
causing any initial clustering in periods to remain.  
Also, Eq.~\ref{eqn:pdecay} predicts that only very compact stars can retain
this clustering for timescales of the order of $10^6~\rm{yrs}$.
This can be understood physically by realizing that,
given a magnetic field strength and period,
a more compact star will have a higher magnetic energy density in each vortex.
This causes each vortex expulsion event to remove greater amounts of magnetic
field from the system, making magnetic braking become increasingly more ineffective.

The observed periods of RQINSs are indeed clustered, however the range
of this clustering and the mean on which it is centered is inconclusive.
Upon detection of more RQINSs, we will be able to conclude more confidently
whether the mean period of RQINSs is higher than that of AXPs and SGRs, and,
this in turn will allow us to predict more accurately the radius
of quark stars.
%%%%%%%%%%%%%%%%%%%%%%%%%%%%%%%%%%%%%%%%%%%%%%%%%%%%%%%%%%%%%%%%%

\section{Conclusion}\label{sec:conclusion}

\begin{figure}
\plotone{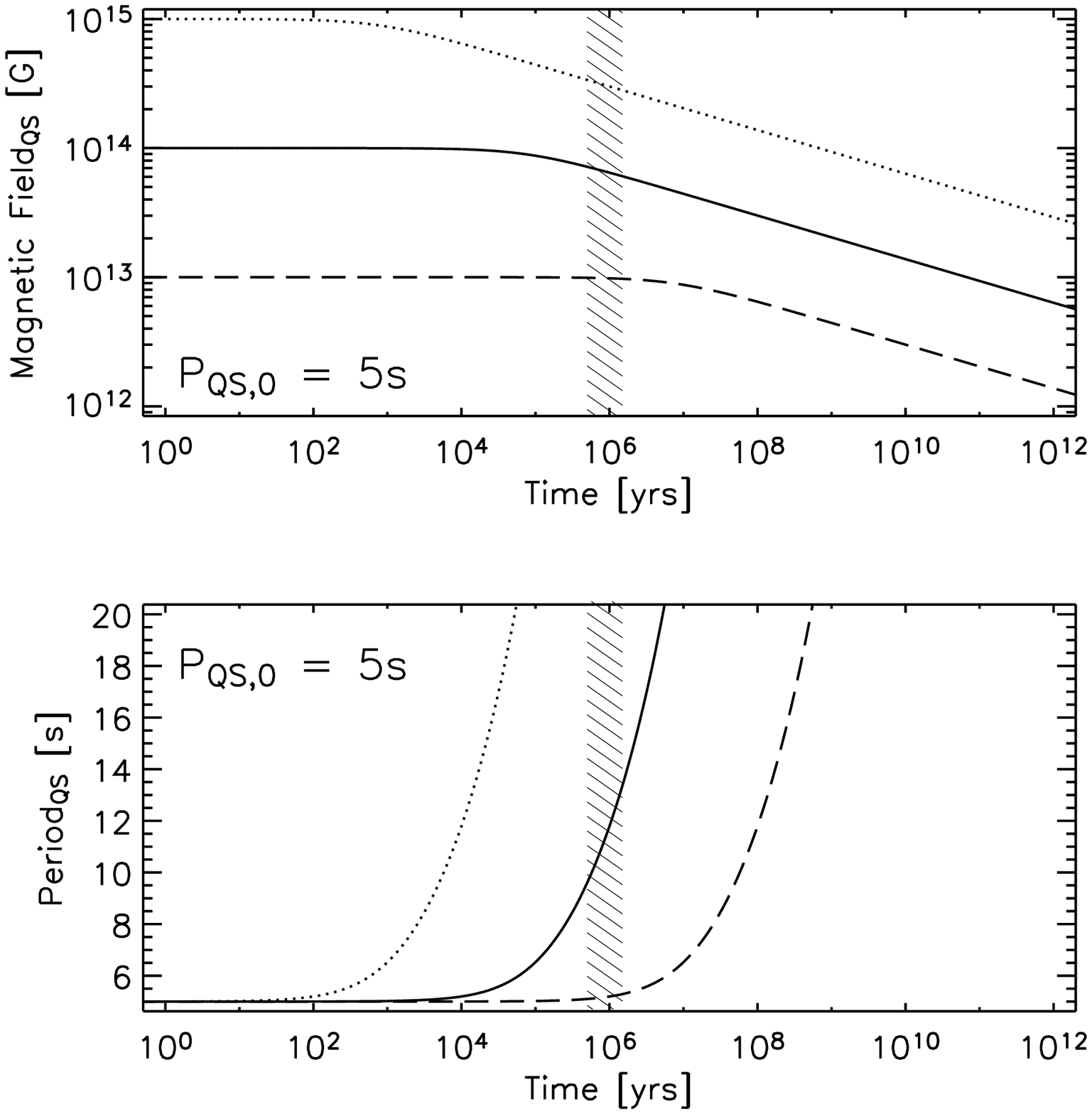}
\caption{\label{fig:b_decay}
Magnetic field decay and spin-down for three different values of initial
field strength ($10^{13}$, $10^{14}$, and $10^{15}~\rm{G}$).
The quark star has an initial period of $5~\rm{s}$
and a radius of $10~\rm{km}$.
The dashed region represents the only two RQINSs having an inferred age.}
\end{figure}

We have shown in this Letter that, if CFL quark stars are born with periods 
and magnetic fields characteristic of AXPs/SGRs, 
then both period and field will remain unchanged within a factor of two 
for timescales of the order of $5\times 10^{5}$ to $5\times 10^{7}~\rm{yrs}$
(Figure~\ref{fig:b_decay}).
Therefore, because AXPs/SGRs are born within a narrow period range, their periods
will remain clustered for timescales applicable to observations. 
Moreover, after timescales of $10^{5}$ to $10^{6}~\rm{yrs}$ 
their periods will be on average higher. 
However, because only a relatively small number of RQINSs have been discovered to date,
it is difficult to determine whether 
their period mean is indeed higher than that of AXPs/SGRs.  

\begin{figure}
\plotone{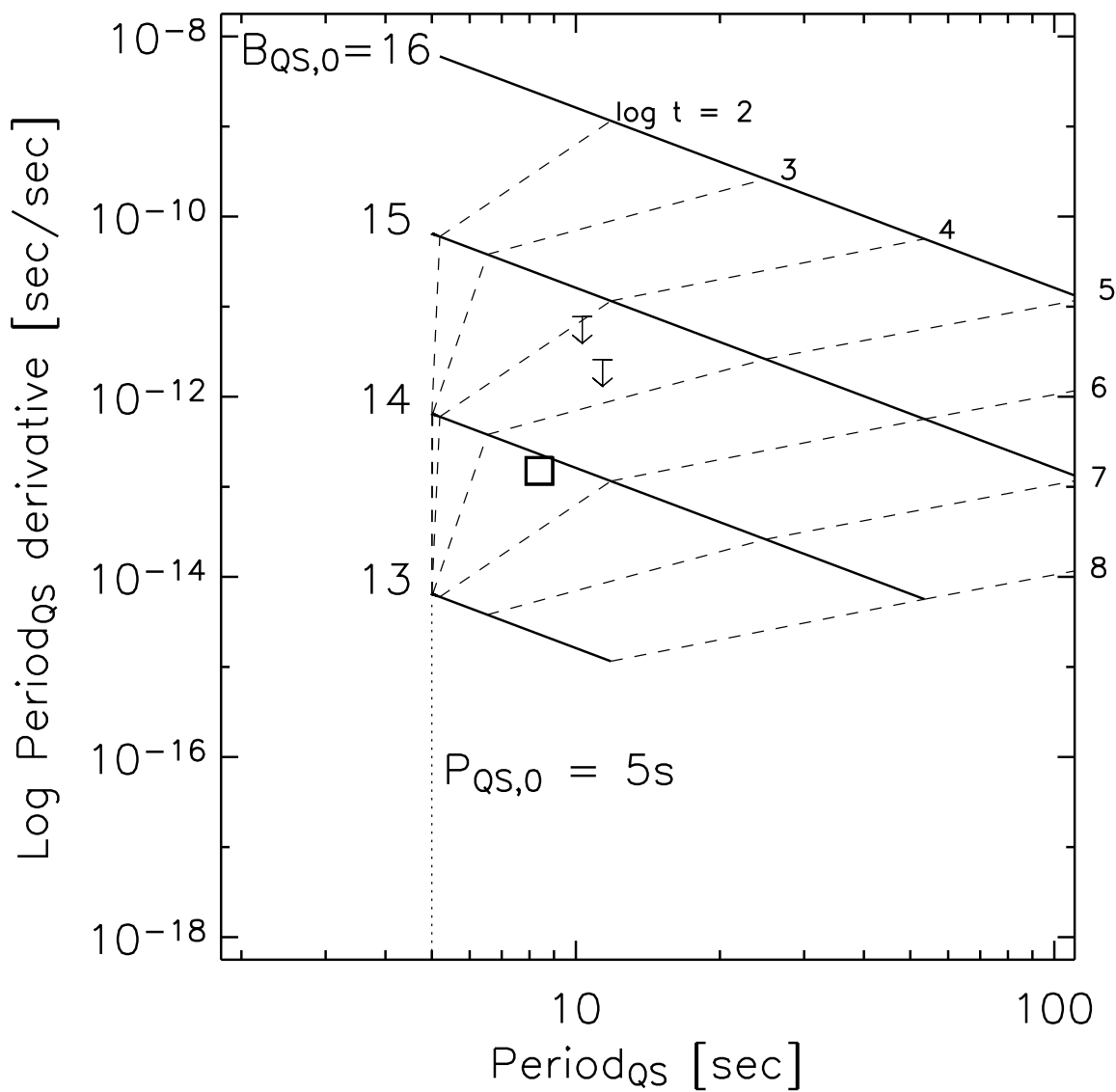}
\caption{\label{fig:p_pdot}
Evolutionary tracks (solid lines) for differing 
surface magnetic fields as indicated, 
and an initial period of $5~\rm{s}$.  
The quark star is assumed to have a
$10~\rm{km}$ radius and a mass of $1~M_{\odot}$. RQINSs
are marked with the small box or down arrows
(for sources with only an upper limit on $\dot{P}$).
Dashed lines represent time in years from the birth of the quark star.
All evolutionary tracks lead to birth parameters
indicative of AXPs/SGRs.}
\end{figure}

Also, within the context of the quark star model, 
we have used a $P-\dot{P}$ diagram (Figure~\ref{fig:p_pdot})
to illustrate how the
field strengths and periods of AXPs/SGRs evolve in a manner
indicative of RQINSs. 
Considering the quark star's magnetic field becomes that of an aligned dipole,
our model provides a natural explanation 
as to why no radio pulsations are observed in RQINSs.

\begin{figure}
\plotone{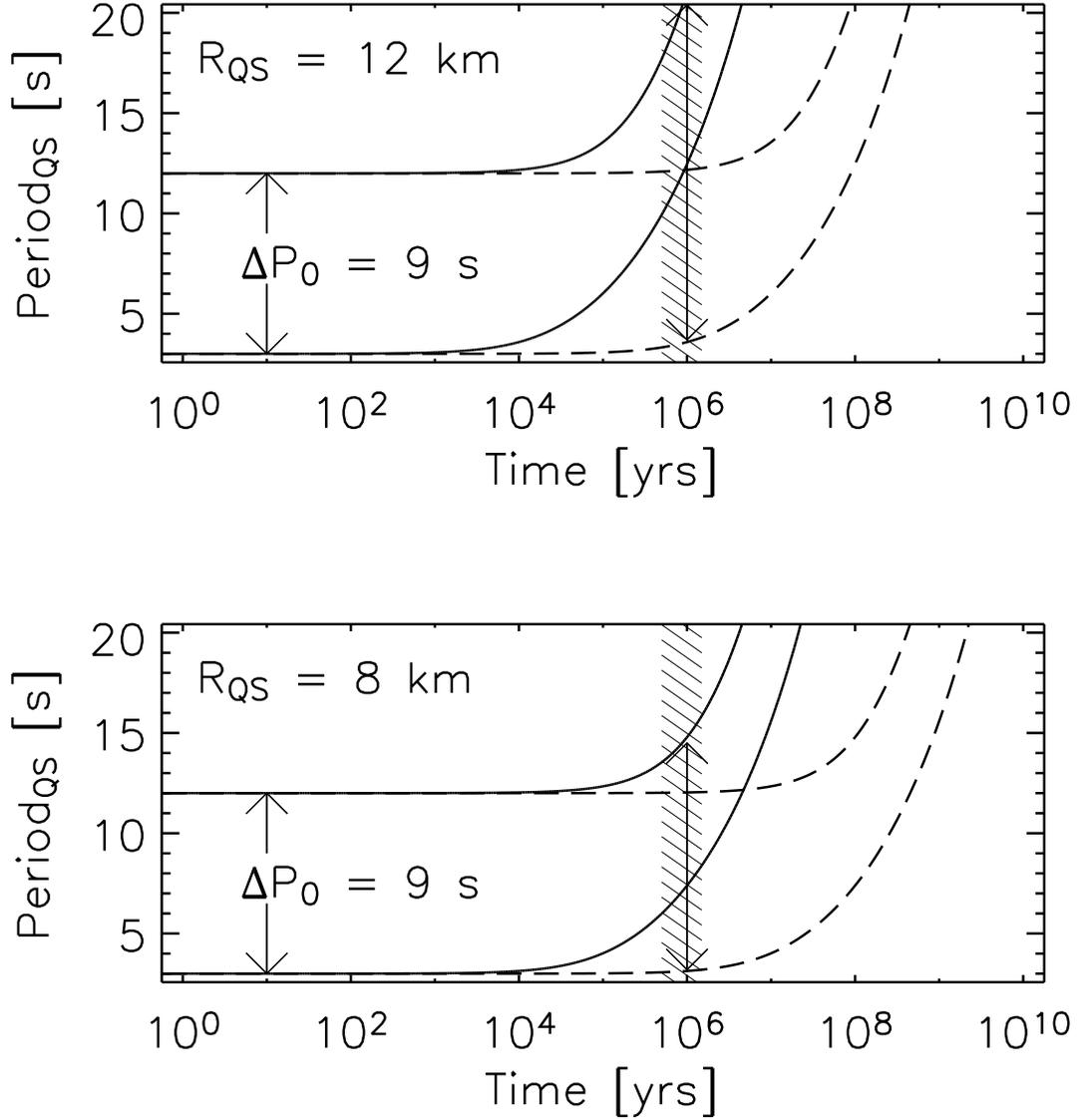}
\caption{\label{fig:p_cluster}
Period clustering evolution for initial magnetic fields
of $10^{13}~\rm{G}$ and $10^{14}~\rm{G}$ (dashed and solid lines respectively),
for radii of $12$ (upper panel) and $8~\rm{km}$ (lower panel).
The upper and lower curves in each panel are the maximum and minimum
of the period range.  If $R_{\rm{QS}} > 10~\rm{km}$, then after $10^6~\rm{yrs}$
the period range becomes large, suggesting only compact objects undergoing
vortex expulsion can explain period clustering in RQINSs.}
\end{figure}

Finally, from Figure~\ref{fig:p_cluster}, we have shown that the
compactness of a quark star is related to period clustering. 
Only highly compact objects are able to maintain clustering in
their periods for timescales of $10^6~\rm{yrs}$, suggesting that
the observed clustering in RQINS periods is an indicator of a 
compact quark star progenitor.  Further detections of RQINSs will allow
us to determine exactly how compact these progenitors are.

%%%%%%%%%%%%%%%%%%%%%%%%%%%%%%%%%%%%%%%%%%%%%%%%%%%%%%%%%%%%%%%%%

\begin{acknowledgements}
We thank K. Mori for his insightful discussions. 
%B.~N. thanks Sigma-Xi for its Grant-in-aid of research.
This research is supported by grants from the Natural Science and
Engineering Research Council of Canada (NSERC).
\end{acknowledgements}

%%%%%%%%%%%%%%%%%%%%%%%%%%%%%%%%%%%%%%%%%%%%%%%%%%%%%%%%%%%%%%%%%

\clearpage

\clearpage

\clearpage

\end{document}